# Collective modes and generation of a new vortex in a trapped Bose gas at finite temperature.

Abdelâali Boudjemâa[1]

*Department of Physics, Faculty of Sciences, Hassiba Benbouali University of Chlef*
*P.O. Box 151. 02000, Chlef, Algeria.*

**Abstract**

The dynamics of Bose-Einstein condensate (BEC) is studied at nonzero temperatures using our variational time-dependent-HFB formalism. We have shown that this approach is an efficient tool to study the expansion and collective excitations of the condensate, the thermal cloud and the anomalous correlation function at nonzero temperatures. We have found that the condensate and the anomalous density have the same breathing oscillations. We have investigated, on the other hand, the behavior of a single quantized vortex in a harmonically trapped BEC at nonzero temperatures. Generalized expressions for vortex excitations, vortex core size and Kelvin modes have been derived. An important and somehow surprising result is that the numerical solution of our equations predicts that the vortex core is partially filled by the thermal atoms at nonzero temperatures. We have shown that the effect of thermal fluctuations is important and it may lead to enhancing the size of the vortex core. The behavior of the singly anomalous vortex has also been studied at nonzero temperatures.

---

[1] Corresponding author :E-mail: a.boudjemaa@univ-chlef.dz

## 1. Introduction

Ultracold Bose gases at nonzero temperatures have recently proven to be a rich field of investigation especially that all experiments actually take place at nonzero temperatures. The effects of finite temperatures are so important, in particular on the thermal cloud, the anomalous density, the expansion of the condensate and on the thermodynamics of the system. Furthermore, the effects of nonzero temperature become mainly obvious in low dimensional systems, where the condensate exhibits fluctuations in its phase. Useful theoretical models have been developed to describe the dynamical behavior of BEC at nonzero temperatures. Among them we can cite generalized mean field treatments [1-4], number-conserving approaches [5-7], classical field theory [8-12], stochastic approaches [13-15] and kinetic approach [16-19].

Alternatively, in this paper we use our TDHFB (time-dependent-Hartee-Fook-Boboliubov) formalism [20-22] which is non-perturbative and non-classicalfield approach. The TDHFB equations are time-dependent variational equations derived using the Balian and Vénéroni (BV) principle [23]. They are a set of coupled time-dependent mean field equations for the condensate, the thermal cloud, and the anomalous average. We have to mention at this point that these equations are quite general and fully consistent as they do not require any simplifying assumptions on the noncondensed or the anomalous densities.

At nonzero temperatures, the dynamic of BECs, such as collective modes and vortices are important sources of information about the nature of the condensate and the thermal cloud. Experimentally, the measurements of these modes can be carried out with high precision with the aim to point out the role of the interactions and quantum correlations [24-26]. Previous theoretical works show that below the transition temperature, the excitations have weak temperature dependence and when the condensate goes to zero the modes approach those of noninteracting trapped gas [27] while they deviate from each other for a large number of particles [27, 28]. It has been shown also that the insertion of the anomalous density in the generalized HFB theory provides a downward shift in the modes observed experimentally near the critical region.

Moreover, the collective modes of the condensate and the thermal cloud have been tested successfully against experiments in the so-called ZNG theory (Zaremba,

Nikuni, and Griffin) [16-18]. In such an approach, the thermal cloud itself is described by a quantum Boltzmann equation coupled to the condensate.

Although these theories give good results against experiments, they completely raised the collective modes of the so-called anomalous density. Certainly this quantity plays a crucial role in Bose gases as well as its absence leads to instabilities in such systems [21, 22]. It is therefore instructive to use our TDHFB formalism within the hydrodynamic approach to study the excitation modes of the anomalous density and its expansion after a sudden switching off of the trap, and this is the subject of the first part of the present paper.

On the other side, many experimental and theoretical efforts have been directed towards the formation and the behavior of vortices in atomic BEC [29-48]. Actually, vortices can be created using a range of different techniques. The development of these techniques has opened a wide door to study more complicated configurations, starting from vortex lattices [30, 31] passing to the creation of a small tangle of vortices [34, 49]. Moreover, vortices in two-dimensional (2D) degenerate Bose gases have also realized [50] such vortices play an important role in the occurrence of the phase transition of the quasicondensate in 2D geometry [51]. Vortex dipoles have also been recently realized experimentally in dilute Bose gas [52-54]. Additional stationary vortex cluster configurations, such as vortex tripoles [55] and other, more exotic arrangements have also been predicted [56].

However, self-consistent but not variational approaches [17] have led to the conclusion that a vortex which is thermodynamically unstable at vanishing temperatures could be stabilized at finite temperature due to the presence of a thermal cloud causing the vortex to dissipate energy and spiral out of the condensate. Alongside this spiraling behavior, the vortex core can become macroscopically occupied by the thermal cloud [37,38,47]. It has also been shown that the thermal cloud density acts as a pinning center and causes the opposite sense of precession which is analogous to the violation of the Kohn theorem in the HFB theory [37,47]. To restore the proper behavior one must treat the dynamics of the thermal cloud in a consistent fashion; this is what our TDHFB theory provides.

Our motivation in the second part of this paper is to revisit the behavior of vortices in Bose gases where we will investigate the effects of temperature on vortex frequencies and the radius of the vortex core by solving analytically and numerically the TDHFB equations. What is advantageous in our theory is that both the thermal cloud and the anomalous density are not considered to be static as in earlier treatments, but are treated dynamically on the same balance as the condensate. This more consistent treatment counteracts the idea that a static thermal cloud can destabilize the vortex [37, 47]. Additionally, our model permits us to go further and predict a new kind of vortices which appear at nonzero temperature namely "anomalous vortices".

The rest of the paper is organized as follows. In Sec. II, we briefly review the derivation of the TDHFB equations. Using the hydrodynamic approach, we show that TDHFB equations satisfy all conservation laws as well being gapless. In Sec. III, we calculate the breathing modes of the anomalous density in the limit of the Thomas-Fermi (TF) approximation. In Sec. IV, we apply our TDHFB formalism to study the behavior of vortices at nonzero temperatures, where we have generalized standard expressions of vortex frequencies and the radius of the vortex core (Sec.IV.A). Next, we compare our results with recent theoretical calculations. The vortex profiles at different ranges of temperatures are also analyzed (Sec.IV.B). In Sec.IV, we shed some light on properties of the so-called anomalous vortex. Our concluding remarks are presented in Sec.V.

## 2. The TDHFB theory

In this section, we briefly discuss the TDHFB equations and the advantages of using such a model before presenting our results. The TDHFB theory based on the Balian-Vénéroni variational principle describes the dynamics of interacting trapped Bose systems at nonzero temperatures. For a short-range interaction potential, the TDHFB equations read

$$i\hbar \dot{\Phi} = \left( -\frac{\hbar^2}{2m}\Delta + V_{ext}(r) + g|\Phi|^2 + 2g\tilde{n} \right)\Phi + g\,\tilde{m}\,\Phi^*, \tag{1.a}$$

$$i\hbar \dot{\tilde{m}} = g(2\tilde{n}+1)\Phi^2 + 4\left( -\frac{\hbar^2}{2m}\Delta + V_{ext}(r) + 2gn + \frac{g}{4}(2\tilde{n}+1) \right)\tilde{m}, \tag{1.b}$$

$$i\hbar \dot{\tilde{n}} = g\left( \tilde{m}^*\,\Phi^2 - \tilde{m}\,\Phi^{*2} \right). \tag{1.c}$$

with $m$ being the atom mass, $V_{ext}(r)$ the external confining potential and $g = 4\pi\hbar^2 a/m$ the coupling constant with $a$ is the s-wave scattering length.

In the set (1), $\Phi$ is the order parameter, $n_c = |\Phi|^2 = |\langle\psi(\vec{r})\rangle|^2$ is the condensate density, $\tilde{n}(\vec{r}) = \langle\psi^+(\vec{r})\psi(\vec{r})\rangle - \langle\psi^+(\vec{r})\rangle\langle\psi(\vec{r})\rangle$ is the thermal cloud, $\tilde{m}(\vec{r}) = \langle\psi(\vec{r})\psi(\vec{r})\rangle - \langle\psi(\vec{r})\rangle\langle\psi(\vec{r})\rangle$ is the anomalous density and $n = n_c + \tilde{n}$ is the total density.

One may understand in few words how Eqs.(1) have been derived simply by recalling that the BV variational principle provides dynamical equations for the variational parameters of the density operator. These parameters are directly related to the previous expectation values (with respect to the density operator) of the operators $\psi(\vec{r})$, $\psi^+(\vec{r})\psi(\vec{r})$ and $\psi(\vec{r})\psi(\vec{r})$, which determine the various densities. For further computational details, see Refs. [20-22]. Moreover the quantities $\tilde{n}$ and $\tilde{m}$ are related by the following equality [23,57]

$$\frac{I-1}{4} = \tilde{n}(\tilde{n}+1) - |\tilde{m}|^2 . \tag{2}$$

If $I \to 1$ or ($T \to 0$), Eq.(2) shows that the absolute value of the anomalous density is larger than the noncondensed density. This proves the importance of the former especially at low temperature, where it cannot be neglected whatever the conditions.

## 2. A. Conservation laws

As known, the anomalous density is a divergent quantity in any geometry. One of the most efficient tools to circumvent this divergence is the renormalization of the coupling constant. Following the method of Burnett *et al* [9, 27], we get from Eq.(1.a)

$$g|\Phi|^2\Phi + g\,\tilde{m}\,\Phi^* = g\left(1 + \frac{\tilde{m}}{\Phi^2}\right)|\Phi|^2\Phi = U|\Phi|^2\Phi . \tag{3}$$

This is similar to the so-called G2 approximation [9, 27] based on the *T*-matrix calculation, which is gapless mean-field theory taking into account effects of the background gas on colliding atoms.

At very low temperature and for dilute gas, $\tilde{m}/\Phi^2 \ll 1$. Therefore, the new coupling constant $U$ reduces immediately to $g$.

Then by introducing $U(r)$ in Eqs.(1), and using the fact that at very low temperature we have from Eq. (2) $2\tilde{n} + 1 \approx 2\tilde{m}$, one obtains

$$i\hbar\dot{\Phi} = \left(-\frac{\hbar^2}{2m}\Delta + V_{ext}(r) + g\left(\beta|\Phi|^2 + 2\tilde{n}\right)\right)\Phi , \tag{4.a}$$

$$i\hbar \dot{\tilde{m}} = \left( -\frac{\hbar^2}{2m}\Delta + V_{ext}(r) + 2g(G\tilde{m} + n) \right)\tilde{m}, \quad (4.b)$$

where $\beta = U/g$ and $G = U/4(U-g)$.

Note that if $\beta = 1$ i.e. $\tilde{m}/\Phi^2 = 0$, Eq.(4.a) reduces to the well-known HFB-Popov equation which is of course safe from all ultraviolet and infrared divergences and thus provides a gapless spectrum.

In a homogeneous system the hydrodynamic excitations are sound waves, while for trapped gas the excitations are not plane waves anymore and have to be classified according to the symmetries present in the trap geometry. Besides the low-lying excitations, which are studied by shaking the gas out of the ground state into the lowest excited states, it is also important to consider time-of-flight experiments, in which the sample is released from the trap, and expands freely in space. Both types of phenomena can be investigated within the hydrodynamic formalism, which we derive now starting from the TDHFB equations.

Hence, a useful reformulation of the set (1) is obtained by factorizing the condensate wave function and the anomalous density according to the Madelung transformation:

$$\Phi(\vec{r},t) = \sqrt{n_c(\vec{r},t)}\, e^{iS(\vec{r},t)}, \quad (5.a)$$

$$\tilde{m}(\vec{r},t) = \sqrt{\tilde{m}(\vec{r},t)}\, e^{i\theta(\vec{r},t)}, \quad (5.b)$$

where $S$ and $\theta$ are phases of the order parameter and the anomalous density respectively. They are real quantities, related to the superfluid and thermal velocities respectively by $v_c = (\hbar/m)\nabla S$ and $v_{\tilde{m}} = (\hbar/m)\nabla \theta$. By substituting expressions (5) in Eqs. (4.a) and (4.b) and separating real and imaginary parts, one gets the following set of hydrodynamic equations:

$$\frac{\partial \sqrt{n_c}}{\partial t} + \nabla(n_c . v_c) = 0, \quad (6.a)$$

$$\frac{\partial \sqrt{\tilde{m}}}{\partial t} + \nabla(\tilde{m} . v_{\tilde{m}}) = 0. \quad (6.b)$$

Equations.(6) are nothing more than equations of continuity expressing the conservation of mass, and Euler-like equations read:

$$m\frac{\partial v_c}{\partial t} - \frac{1}{2}m v_c^2 = -\nabla \left[ -\frac{\hbar^2}{2m}\frac{\Delta\sqrt{n_c}}{\sqrt{n_c}} + V_{ext} + g(\beta n_c + 2\tilde{n}) \right], \quad (7.a)$$

$$m\frac{\partial v_{\tilde{m}}}{\partial t}-\frac{1}{2}mv_{\tilde{m}}^{2}=-\nabla\left[-\frac{\hbar^{2}}{2m}\frac{\Delta\sqrt{\tilde{m}}}{\sqrt{\tilde{m}}}+V_{ext}+2g(G\tilde{m}+n)\right],\qquad(7.b)$$

where $\left(-\hbar^{2}/2m\right)\Delta\sqrt{n_{c}}/\sqrt{n_{c}}$ and $\left(-\hbar^{2}/2m\right)\Delta\sqrt{\tilde{m}}/\sqrt{\tilde{m}}$ are, respectively, quantum and anomalous pressures.

In a non-stationary situation, it is then considered small oscillations (low density) for the condensed and anomalous densities around their static solutions in the form:

$$\begin{aligned}n_{c}&=n_{c0}+\delta n_{c}\\ \tilde{m}&=\tilde{m}_{0}+\delta\tilde{m}\end{aligned},\qquad(8)$$

where $\delta n_{c}/n_{c0}\ll 1$ and $\delta\tilde{m}/\tilde{m}_{0}\ll 1$.

Shifting the phases by $-\mu_{c}t/\hbar$ and $-\mu_{\tilde{m}}t/\hbar$, we then linearize Eqs.(6) and (7) with respect to $\delta n_{c}$, $\delta\tilde{m}$, $\nabla S$ and $\nabla\theta$ around the stationary solution. The zero order terms give two expressions for the chemical potential:

$$\mu_{c}=-\frac{\hbar^{2}}{2m}\frac{\Delta\sqrt{n_{c0}}}{\sqrt{n_{c0}}}+V_{ext}+g(\beta n_{c0}+2\tilde{n}),\qquad(9.a)$$

$$\mu_{\tilde{m}}=-\frac{\hbar^{2}}{2m}\frac{\Delta\sqrt{\tilde{m}_{0}}}{\sqrt{\tilde{m}_{0}}}+V_{ext}+2g(G\tilde{m}_{0}+n),\qquad(9.b)$$

where $\mu_{c}$ is the chemical potential of the condensate and $\mu_{\tilde{m}}$ is the chemical potential associated with the anomalous density. Strictly speaking $\mu_{\tilde{m}}$ is also associated with the thermal cloud density since $\tilde{n}$ and $\tilde{m}$ are related to each other by Eq.(2).

Clearly $\mu_{c}\neq\mu_{\tilde{m}}$ at all ranges of temperature except near the transition where $n_{c}=\tilde{m}=0$ and $\tilde{n}=n$. Additionally, in the grand canonical ensemble the Hamiltonian may be written as $K=H-\mu N$. If in the experiment only the total number of particles $N=N_{c}+\tilde{N}$ or the total density $n$ can be fixed, then the total chemical potential of the system can be given as

$$\mu=\frac{N_{c}}{N}\mu_{c}+\frac{\tilde{N}}{N}\mu_{\tilde{m}},\qquad(10)$$

where $N_{c}/N$ and $\tilde{N}/N$ are, respectively, the condensed and the thermal fractions. It should be noted that this equation arises naturally from our formalism without any subsidiary assumptions. Moreover, Eq.(10) very nicely guarantees the conservation of the total number of particles and highly coincides with the theory of Ref [58].

After the above analysis we can confirm that the TDHFB equations satisfy all the conservation laws such as the energy and the total number of particles. Additionally, they are characterized by a gapless excitation spectrum, which is compatible with the finite temperature version of the Hugenholtz-Pines theorem [59, 60].

Let us now consider a harmonic oscillator potential ($V_{ext}(r) = m\omega_0^2 r^2 / 2$) with a large number of particles. It is legitimate in this situation to neglect the kinetic energy associated with both quantum and anomalous pressures. Therefore, Eqs.(9) provide useful formulas for the radius of the condensate and the anomalous density, respectively, as

$$\frac{R_{TF}^5}{N_c} = C\left(\frac{2}{\frac{N_c}{N}} + \beta - 2\right), \qquad (11.a)$$

$$\frac{R_{TF}^{\tilde{m}\,5}}{\tilde{M}} = C\left(\frac{1}{\frac{\tilde{M}}{N}} + G\right), \qquad (11.b)$$

where $C = R_{TF}^{(0)5}/N$ with $R_{TF}^{(0)} = a_0(15Na/a_{H0})^{1/5}$ is the standard TF approximation radius at zero temperature and $a_{H0} = \sqrt{\hbar/m\omega_0}$ is the harmonic oscillator length. $\tilde{M}/N$ is the anomalous fraction where $\tilde{M} = \int d\vec{r}\,\tilde{m}(r)$ is the integrated value of the anomalous density [22].

The relation (11.a) reproduces the overall behavior observed experimentally in [61] as well as yielding the zero temperature expression for $N_c/N = \beta = 1$. Nevertheless, as can be seen from Fig.1, $R_{TF}^5/N_c$ increases with increasing $N_c/N$ and gives reasonable agreement with both theoretical treatments of HFB-Popov and experimental results of [61] for small values of $\beta$.

Furthermore, despite the lack of experimental data of the anomalous density in the literature, we can point out from expression (11.b) that the radius of the anomalous density is small compared to that of the condensate at low temperature. At high temperature both radii should vanish since $n_c = \tilde{m} = 0$ [21].

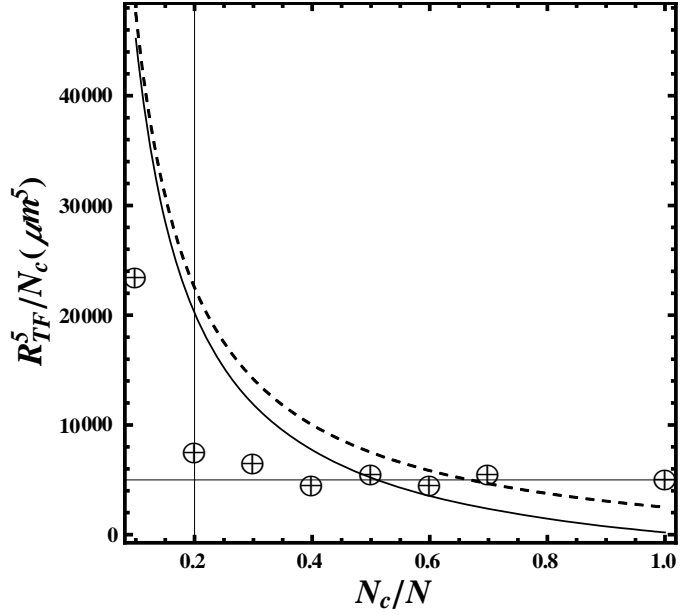

**FIG. 1.** The ratio $R_{TF}^5/N_c$ as function of the condensed fraction. Circles show experimental results of [62], dashed line: HFB-Popov calculations ($\beta=1$) and solid line is our predictions with $\beta=0.08$.

### 3. Breathing modes of the anomalous density

As an application of our implementation of the TDHFB equations, we study the breathing oscillation of a BEC at nonzero temperatures.

Inserting Eqs.(9) into (7) and taking the time derivative of the resulting equations, one finds

$$m\frac{\partial^2 \delta n_{c0}}{\partial t^2} = \nabla(n_{c0}\nabla\delta\mu_c), \tag{12.a}$$

$$m\frac{\partial^2 \delta \tilde{m}_0}{\partial t^2} = \nabla(\tilde{m}_0\nabla\delta\mu_{\tilde{m}}). \tag{12.b}$$

Equations.(12) describe the collective modes of both condensate and anomalous density for Bose gas in an arbitrary potential. So they form in this sense a natural extension of the famous Stringari equation [62]. It is to be noted that similar equations have been derived within the ZNG theory [16] but without taking into account the anomalous density.

The calculation of the collective modes in a trapped case is not trivial at nonzero temperature due to the fast extent of the cloud and the spatial variation of the coherence length. In the spirit of the TF approximation, it is therefore necessary to explore the properties of the collective modes when both pressures are neglected from the equations of motion.

Before proceeding further, it is important to note that the kinetic term of the thermal cloud does not appear explicitly in the equations but is rather hidden in Eq. (1.c). Indeed, the kinetic term of the thermal cloud is related to the second derivative of the anomalous density. Differentiating Eq.(2) yields a relation of the form: $\Delta \tilde{n} \approx (\nabla \tilde{m})^2 - (\nabla \tilde{n})^2 + \tilde{m}\Delta\tilde{m}$, which shows clearly that neglecting $\Delta \tilde{m}$ does not necessarily mean neglecting $\Delta \tilde{n}$ and therefore omitting the anomalous pressure does not mean neglecting the thermal pressure [63]. Such feature we shall adopt in what follow.

When the anomalous pressure is neglected, Eq.(9.b) reduces to $\bar{\mu}_{\tilde{m}} = \mu_{\tilde{m}} - 2gn = V_{ext} + 2gG\tilde{m}$, since the total density is conserved ($\delta n = 0$). Thus

$$\delta\bar{\mu}_{\tilde{m}} = 2gG\delta\tilde{m}. \tag{13}$$

The anomalous density becomes

$$\tilde{m}_0 = \frac{\bar{\mu}_{\tilde{m}} - V_{ext}}{2gG}. \tag{14}$$

Introducing Eqs.(13) and (14) into (12.b) one finds

$$\frac{\partial^2 \delta\tilde{m}_0}{\partial t^2} = \frac{2gG}{m}\nabla(\tilde{m}_0 \nabla \delta\tilde{m}_0). \tag{15}$$

In the TF approximation the chemical potential and the radius of the anomalous density are related by $\bar{\mu}_{\tilde{m}} = m\omega_0^2 R_{TF}^{\tilde{m}\,2}/2$. Assuming oscillations with time dependence $\delta\tilde{m}_0 \propto e^{-i\omega t}$, and working in the spherical coordinates, the differential equation (15) simplifies to

$$2\Omega^2 \chi_l(z) = 2y\frac{d\chi_l(z)}{dz} + (1-z^2)\left[\frac{d^2}{dz^2} + \frac{2}{z}\frac{d}{dz} - \frac{l(l+1)}{z^2}\right]\chi_l(z), \tag{16}$$

where $\chi_l(y) = \delta\tilde{m}_0 / Y_l^m(\theta,\varphi)$, $\Omega = \omega/\omega_0$ and $z = r/R_{TF}^{\tilde{m}}$.

In terms of the dimensionless coordinate $x = z^2$, Eq. (16) will be valid for $0 \leq x \leq 1$ and hence it takes the standard form of the hypergeometric function $F(\alpha,\beta,\gamma;x)$. For low-energy excitations with orbital angular momentum $l = 0$, one can obtain after a little algebra values of the excitation energy

$$\Omega_j = \sqrt{j(2j+3)}. \tag{17}$$

For $j = 1$ we get a surprising result $\omega = \sqrt{5}\omega_0$, i.e. we recover the breathing mode obtained earlier for the condensate. This shows that the condensate and the anomalous density dilate and contract together at the same time and with the same frequency, which constitutes a new feature for ultracold Bose gases at finite temperature. It is important to mention here that we are able to study the evolution of the anomalous density when the trap is switched off suddenly by extending the TF approximation Eq.(15) in the time-dependent harmonic potential. The primary result shows that the anomalous density in the TF regime keeps its shape at any moment. Analogous result was found by Castin and Dum [64] for the condensate.

## 4. Vortices at nonzero temperatures

### 4. A. Vortex frequencies

Consider a straight vortex line in a BEC in the trapping geometry of an ideal cylinder. The $z$-direction is free, and in the $x$, $y$-plane one has a harmonic confining potential $V_{ext}(r) = m\omega_r^2 r^2 / 2$, where $r^2 = x^2 + y^2$. We will try to find the eigenfrequency and wavefunction of an excitation corresponding to the rotation of the vortex line around the $z$-axis. Due to the instability of multiquantum vortices [37,38,65,66], we will focus on looking for the solution of the stationary TDHFB equation (4.a) with orbital angular momentum 1, we then obtain:

$$i\hbar\dot{\Phi} = -\frac{\hbar^2}{2m}\left(\frac{d^2}{dr^2} + \frac{1}{r}\frac{d}{dr} - \frac{1}{r^2}\right)\Phi + \left[V_{ext}(r) + g\left(\beta|\Phi|^2 + 2\tilde{n}\right)\right]\Phi, \qquad (18)$$

In the TF limit, the radius of the trapped BEC in the $x$, $y$-plane is $R_{TF} = \sqrt{2\mu_c/m\omega_r^2}$. The chemical potential is $\mu_c \approx gn_{TF}(0)$, with $n_{TF}(0) \approx n_c(0)$ being the finite temperature TF density at the center of the trap. Such estimation can be attributed to the fact that the thermal atoms are usually localized at the edge of the trap where they develop a peak [2,20,43]. Therefore, in the center of the trap the noncondensed density should vanish ($\tilde{n}(0) \to 0$). This behavior is also valid for the anomalous density [21].

Assuming now that $R_{TF} \ll \xi$ where

$$\xi = \frac{\hbar}{\sqrt{m\mu_c}} = \frac{\xi^{(0)}}{\sqrt{n_c/n}}, \qquad (19)$$

is an estimate vortex size at finite temperature and $\xi^{(0)} = \hbar/\sqrt{mng}$ is the standard vortex size at zero temperature.

In this case one can write an approximate solution of Eq. (18) as

$$\Phi(r) = \Phi_{TF}(r) f(r/\xi) e^{i\phi}, \qquad (20)$$

where $\Phi_{TF}(r)$ is the TF wavefunction (see below) and $R_{TF}$ is the finite temperature TF radius.

For a large condensate, it is natural then to write $\Phi_{TF}(r)$ via the expression (9.a) as

$$\Phi_{TF}(r) = \sqrt{\frac{\mu_c}{g}\left(1 - \frac{r^2}{R_{TF}^2}\right)}. \qquad (21)$$

Indeed, we may easily show that upon linearizing Eq. (18) around a static solution by using the parametrization $\Phi = \Phi_0 + \sum_k \left(u_k e^{-i\omega_k t} - v_k e^{i\omega_k t}\right)$ in which $\omega_k$ are the quasi-particle frequencies and $u_k$ and $v_k$ are the quasi-particle amplitudes, we get trivially the Bogoliubov-de Gennes (BdG) equations [20]. The resulting equations cannot be solved exactly. Luckily in many cases, one can use the local density approximation. In the spirit of this approximation, we write $u_k(r) = \bar{u}$ and $v_k(r) = \bar{v} e^{-2i\phi}$ [67] and set $\omega_k(r) = \omega_v$. Therefore, the BdG equations for these functions read:

$$\hbar\omega_v \bar{u} = \left[-\frac{\hbar^2}{2m}\left(\frac{d^2}{dr^2} + \frac{1}{r}\frac{d}{dr}\right) + V_{ext} + 2g\left(\beta|\Phi|^2 + \tilde{n}\right) - \mu_c\right]\bar{u} - g\beta\Phi^2 \bar{v}$$

$$-\hbar\omega_v \bar{v} = \left[-\frac{\hbar^2}{2m}\left(\frac{d^2}{dr^2} + \frac{1}{r}\frac{d}{dr} - \frac{4}{r^2}\right) + V_{ext} + 2g\left(\beta|\Phi|^2 + \tilde{n}\right) - \mu_c\right]\bar{v} - g\beta\Phi^2 \bar{u} \qquad (22)$$

We now assume that the solutions of Eqs. (22) are given by:

$$\bar{u} = \frac{1}{\sqrt{4\pi L}}\left(\frac{f}{r} + \frac{\partial f}{\partial r}\right)\Phi_{TF}$$

$$\bar{v} = \frac{1}{\sqrt{4\pi L}}\left(\frac{f}{r} - \frac{\partial f}{\partial r}\right)\Phi_{TF} \qquad (23)$$

where $L$ is the length of the vessel.

Next we introduce Eqs. (18), (21) and (23) into the set (22). After that we multiply the sum of the two resulting equations by $(\bar{u} + \bar{v})$ and integrate over $d^3r$. This yields

$$\hbar\omega_v = \frac{2\hbar^2}{m}\int\left\{\frac{f^2}{r^2}\left(\frac{d^2\Phi_{TF}}{dr^2} + \frac{1}{r}\frac{d\Phi_{TF}}{dr} + \frac{2}{f}\frac{d\Phi_{TF}}{dr}\frac{df}{dr}\right)\right\}\frac{d^3r}{4\pi L}. \qquad (24)$$

The main contribution to the integral in the left-hand-side of Eq. (24) comes from distances where $\xi \ll r \ll R_{TF}$. We then set $f \approx 1$ and $\Phi''_{TF} = (1/r)\Phi'_{TF} = 1/R_{TF}^2$, we obtain

$$\hbar\omega_v = -\frac{2\hbar^2}{mR_{TF}^2}\int_\xi^{R_{TF}}\frac{dr}{r} = -\frac{2\hbar^2}{mR_{TF}^2}\ln\left(\frac{R_{TF}}{\xi}\right). \quad (25)$$

Using the fact that $\mu_c = m\omega_r^2 R_{TF}^2/2$, we get straightforwardly the finite temperature corrections of the vortex frequency as

$$\frac{\omega_v}{\omega_r} = -\frac{2}{(R_{TF}/a_0)^2}\ln\left(\frac{R_{TF}}{\xi}\right), \quad (26)$$

where $R_{TF}$ and $\xi$ are the extended radius and vortex size given respectively by Eqs. (11.a) and (19).

Therefore, the obtained eigenfrequency is negative. This indicates the presence of thermodynamic (energetic) instability as one may expect in a non-rotating trap where the vortex state is not the ground state.

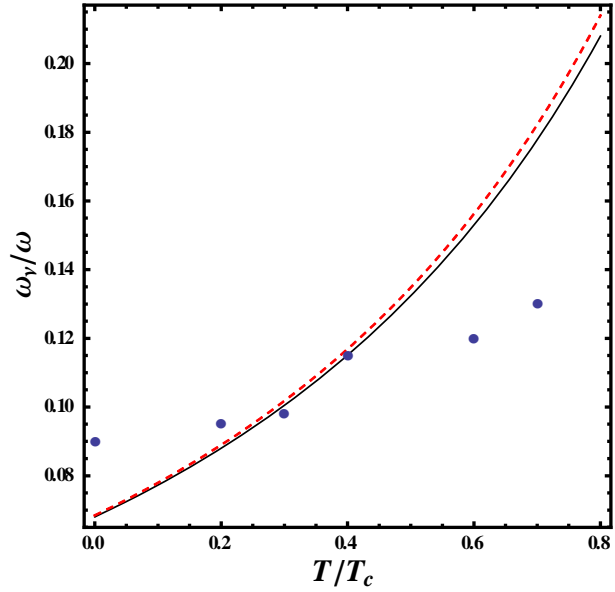

**FIG. 2. (Color online) Vortex frequency as function of reduced temperature for $R_{TF}^{(0)}/\xi^{(0)} = 0.35$ and $\beta = 1.025$. Solid line: our predictions, Red dashed: HFB-Popov ($\beta = 1$) and blue circles: the ZNG calculation [68].**

From Fig.2 we can see that our prediction of Eq.(26) agrees reasonably well with the calculations of the HFB-Popov and ZNG theories [68] at temperatures less than $T = 0.5T_c$ (here we have followed the method outlined in [2,68] to calculate the reduced temperature). At $T \geq 0.6T_c$ our results start to deviate from those of preceding theories.

**4.B. Numerical results**:

To complete the picture, we restrict ourselves in this section to analyze profiles of singly quantized and anomalous vortices at nonzero temperatures by explicitly solving our TDHFB equations.

First of all, we try to see how the singly quantized vortex is generated in the condensed phase and how the thermal part of the system looks? We then have to deal with solving numerically our Eqs. (18), (1.b) and (2). For single vortex lines, cylindrical symmetry is often deployed to reduce the computational cost of numerical solutions [37]. Employing the cylindrical symmetry, the TDHFB equations can be reduced to radial equations, which we discretize using a finite-difference method. The physical parameter values for the gas and the trap have been chosen to be the same as in Refs [54, 68].

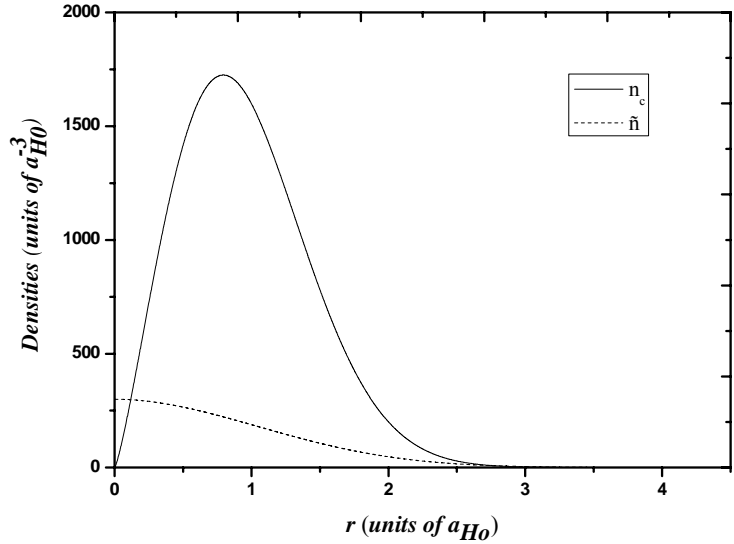

**FIG. 3. Condensate and thermal gas density profiles for** $N/N_c = 75\%$.

At first sight, the vortex core, such as the one seen in Fig.3, appears partially occupied (~10%) by thermal atoms. This is indeed due to the much lower density at the vortex core and, hence, the lower energy cost of gathering particles at that position. However, our results are in qualitative agreement with those obtained in [47, 68-71], where it has been shown that the thermal atoms feel the condensate density as an extra potential and therefore can be located inside the vortex core [69,70]. Additionally, the inclusion of anomalous density, which quantifies correlations of pairs of noncondensed atoms with pairs of condensed atoms, may play a crucial role on the formation and on the shape of vortices at low and intermediate temperatures. Note that if that anomalous correlations are absent, the superfluidity does not occur [21, 22] and hence vortices cannot survive in Bose gas. Thus, the correct description of vortices necessarily requires taking into account uncondensed particles as well as the anomalous density. This is especially important at fast rotation that increases the system energy and by this depletes the condensate producing more uncondensed atoms.

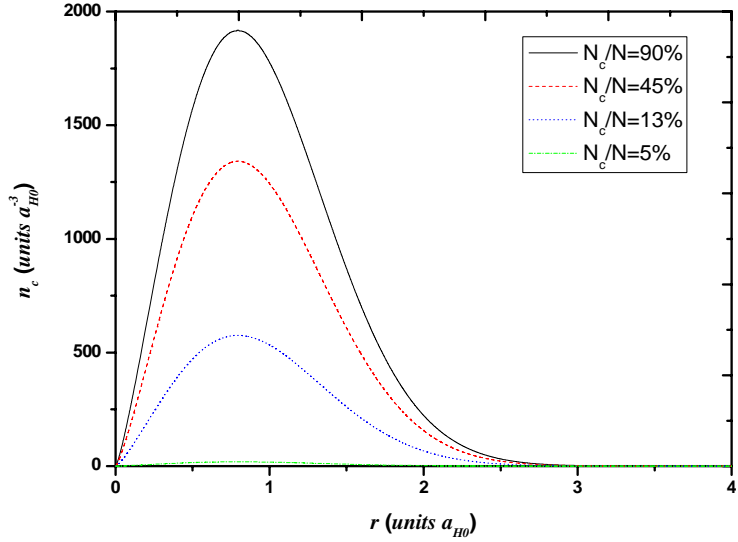

**FIG. 4. (Color online) Condensate density as function of the radial distance for various condensed fractions**

Clearly, we observe from Fig.4 that by decreasing $N_c/N$, the condensed density begins to decrease and starts to disappear when $N_c/N$ approaches 5%. This overall behavior coincides very well with what was obtained earlier in the literature. In addition, the vortex state is shown to be locally stable at all ranges of temperature.

Figure.4 also depicts that the vortex core becomes effectively larger with increasing temperature, and therefore pushed slightly away from the center of the trap in good agreement with both recent Bogoliubov calculations of [71] and with our analytical predictions of Eq.(19) as can be seen in Fig.5. It is now clear that the normal and anomalous correlations, which we have consistently taken into account in our theory, may lead to a large vortex core. These correlations tend also to lower the energy compared to that of the mean field ground state [71].

Experimentally the radius of a vortex core is ordinarily several times smaller than the wavelength of light used for imaging, making direct, *in situ* observation of vortices in a trapped condensate difficult [31,54].

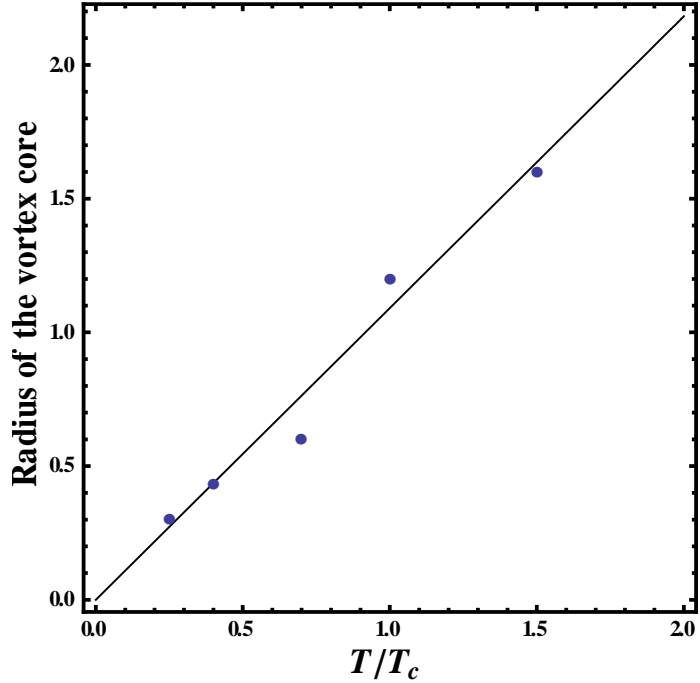

**FIG.5. (Color online) Radius of the vortex core as function of the reduced temperature.**

Indeed, the vortex contrast decreases because the vortex line undergoes Kelvin oscillations (kelvons) [72, 73] due to the presence of thermal fluctuations. In fact, this is true irrespective of the presence or not of the noncondensed atoms in the vortex core. In this case, the Kelvin modes can be calculated easily from Eqs. (22) as $K = E_k \ln(1/k\xi)$ where $\xi$ is the nonzero temperature vortex size defined in Eq.(19). Very recently, the Kelvin collective mode has been determined for rotating BEC containing up to 19 singly quantized vortex filaments, using the microscopic Bogoliubov–de Gennes theory [74].

Finally we extend our study to examine the behavior of the so-called anomalous vortex (associated with the anomalous density). Including then the complex function (5.b) into Eq.(4.b) without imposing the singly quantized vortex on the condensed phase. The resulting equation contains a centrifugal potential which forces the solution of $\tilde{m}$ to be zero along the $z$-axis for nonzero angular momentum.

Again, we solve numerically our TDHFB equations for single vortex with the same experimental values corresponding to Fig.3. In Fig.6 we plot qualitatively the anomalous vortex as a function of temperature. It is easy to see that this type of vortex preserves the same shape as the ordinary vortex whatever the position. The formation of the anomalous vortex occurs first due to the centrifugal forces on the gas and second owing to the correlations between condensed and noncondensed atoms. It arises and grows at low temperature until it reaches its maximum value at intermediate temperatures. After that, it starts to disappear near the transition. This ultimately conducts us to confirm that the anomalous vortex accompanies in analogous manner the ordinary vortex.

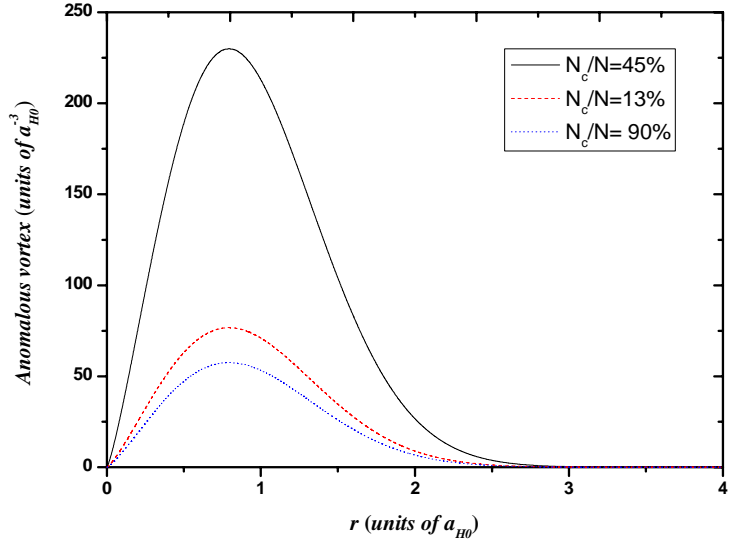

**FIG.6. (Color online) Anomalous vortex vs. the radial distance for various condensed fraction.**

To our knowledge, anomalous vortices have never been investigated in the literature. It is worth noticing, that formulas for vortex frequencies, the radius of the vortex core and Kelvin modes of the anomalous vortex can be derived following the same fashion as in Sec.VI. A.

## 5. Conclusion

Our work is divided into two parts. In the first part, by applying our TDHFB formalism within the hydrodynamic approach, we derived a set of two equations treating self-consistently the expansion and the collective modes of both the condensate and the anomalous density in trapped Bose gas at finite temperature. The main message emerging from our analysis is that at low temperature, the breathing modes of the anomalous density have the same value found earlier for the condensate. Also, the conclusion that we reached in this work shows that the anomalous density in the Thomas-Fermi regime keeps its shape at any moment after a sudden switching off of the trap.

In the second part, we have discussed the effects of the anomalous correlation function and temperature on the properties of vortices in harmonically-trapped Bose gas. In such study, we have generalized in particular standard expressions of the vortex excitations and the size of the vortex core at nonzero temperatures. Our analytical predictions constitute good agreement with ZNG-simulations and HFB-Popov calculations. Moreover, we have explored numerically the full static TDHFB equations in the presence of a single quantized and anomalous vortex. The outcomes of this simulation are numerous. First of all, regarding the quantum vortex, an important and somehow surprising result, is that the TDHFB formalism predicts that the vortex core is partially occupied by the thermal atoms at nonzero temperatures. At this stage, it should be noted that the filling of the vortex core by the thermal cloud is not yet observed experimentally and remains challenging for the experimentalists. Secondly, the size of the core swells with increasing temperature in excellent agreement with both our analytical calculations and recent theoretical predictions [72]. Indeed, the vortex contrast decreases for the reason that the vortex line undergoes Kelvin oscillations. In addition, an extended formula of such Kelvin modes at nonzero temperatures has also been derived in this paper. Furthermore, we have shown that normal and anomalous correlations may lead to modifying the size of the vortex core.

On the other hand, we have investigated the formation and the behavior of the singly anomalous vortex. We have found that this later preserves the same shape as the ordinary vortex. The anomalous vortex reaches its maximum value at intermediate temperatures while it disappears near the transition when the gas becomes completely thermalized.

It should be noted that a doubly quantized vortex can be generated self consistently in the anomalous density if we insert condensed and anomalous phases simultaneously in the TDHFB equations which is in fact an advantage of our formalism. Certainly, further experimental and theoretical effort is required to gain more insight into what indeed is happening about this type of vortex.

An interesting future work is to investigate the properties of quantum and thermal vortices in three and two-dimensional BEC with dipole-dipole interactions at nonzero temperatures.

**Acknowledgments**

The author acknowledges Gora Shlyapnikov, Yuri Kivshar, Jean Dalibard and Eric Cornell for valuable discussions.